\documentclass[11pt,a4paper]{article}
\usepackage{jheppub}

\makeatletter
\gdef\@fpheader{}  
\makeatother

\usepackage{amsmath,amsthm,amssymb,graphics,epsfig,color,mathrsfs,cases,colortbl,framed}
\usepackage{ulem}
\usepackage{graphicx}
\usepackage{subfigure}
\usepackage{booktabs}    
\usepackage{multirow}
\usepackage{tabularx}
\usepackage{longtable}
\usepackage{array}
\usepackage{nicematrix}
\usepackage{caption}

\title{Massive $U(1)$ gauge field and their accompanying scalars in brane world  }

\author{Ye-Hao Yang,}
\author{Wen-Xuan Ma,}
\author{Chun-E Fu}

\affiliation{
Institute of Theoretical Physics, School of Physics, Xi'an Jiaotong University, \\
No.28 West Xianning Road, Xi'an, Shaanxi, China
}
\emailAdd{fuche13@mail.xjtu.edu.cn}
\footnotetext[1]{Corresponding author.}

\abstract{In brane-world scenarios, the effective action of a massless bulk \(U(1)\) gauge field preserves gauge invariance via couplings between massive vector Kaluza-Klein (KK) modes and scalar KK modes. In this work, we extend this framework by introducing a term \((\nabla^M X_M)^2\) into the massless bulk \(U(1)\) gauge action. This modification explicitly breaks the full gauge redundancy while preserving a residual gauge symmetry both in the bulk and on the brane. In this setup, the scalar KK modes can acquire masses from the background geometry. Notably, we find that on the 5D brane, these scalar KK modes are lighter than the vector KK modes. In contrast, on the 6D brane, two types of scalar modes emerge; the mixed interactions between them give rise to oscillations among these scalar modes.
}


\begin{document}

\maketitle
\flushbottom

\section{Introduction}

The idea that our Universe may contain additional spatial dimensions beyond the four-dimensional spacetime has drawn increasing attention over the past decades. The earliest attempt in this direction can be traced back to Kaluza–Klein (KK) theory, which proposed that introducing a fifth spatial dimension enables a unified description of gravity and electromagnetism \cite{KLEIN1926,Maki1962,Cabibbo1963}. This idea was later extended dramatically in string theory, where the presence of multiple extra dimensions is not optional but required for mathematical consistency. Building on these developments, modern brane-world models have provided further insights into the hierarchy problem and the cosmological constant problem. Representative examples include the ADD model \cite{ArkaniHamed1999}, which assumes large but compact extra dimensions, and the Randall–Sundrum (RS) scenarios \cite{Randall1999a,Randall1999}, where warped extra dimensions give rise to novel gravitational dynamics. 

Within the brane-world framework \cite{Rubakov1983a,Rubakov1983,Kobayashi2002,Wang2002,Melfo2003,Dzhunushaliev2008,Brihaye2012,Bazeia2014,Csaki2016,Kanti:2018}, one particularly compelling line of research concerns the behavior of higher-dimensional matter fields and their associated Kaluza–Klein (KK)  modes \cite{Csaki2000,Oda2000,RandjbarDaemi2000,Csaki2003,Melfo2006,Liu2007,Liu2008,Zhao2011,Costa2013,Arai:2017,Guo:2023:2024}. Investigating these KK spectra is crucial not only for ensuring the theoretical consistency of higher-dimensional models but also for probing potential signatures of extra dimensions in observable physics. A growing body of literature has explored how KK excitations may lead to distinctive phenomenology, including modifications to gravitational interactions \cite{Dvali2000,Xie2013,Tan:2025}, corrections to Standard Model processes \cite{Boos2015,Kulaxizi2014}, and potential resonant modes localized near the brane \cite{Liu2009,Guo2015,Jia:2025}. These works highlight the importance of KK modes as a window into the possible structure of higher-dimensional spacetime \cite{Csaki:2018,Malek:2020,Chivukula:2021,Ashmore:2021}.

It is well established that massless fields propagating in the bulk give rise to massive KK modes in the effective four-dimensional theory on the brane, with the masses induced by the geometry of the extra dimensions. For a bulk U(1) gauge field, the KK tower on the brane includes both vector and scalar modes. Notably, the effective four-dimensional action preserves gauge invariance due to vector–scalar and scalar–scalar couplings among the KK modes \cite{Fu2018,Fu2020,Fu2022}. As a consequence, sufficient gauge freedom remains in the brane theory to eliminate the unphysical polarization degrees of freedom from the massive vector KK modes. This mechanism further reveals that the massive scalar KK modes play a role analogous to Stückelberg or Goldstone fields.

Previous studies have indicated that certain mixed vector–scalar and scalar–scalar couplings may exist. However, since the basis functions for the scalar KK modes cannot be uniquely determined, the precise form of these couplings remains uncertain. What is important is that, due to such mixings, the scalar KK modes cannot be fully gauged away—implying that they may carry physical degrees of freedom \cite{Fu2025}. To properly define these scalar modes, and motivated by the Stückelberg mechanism, we propose adding a covariant term \((\nabla^M X_M)^2\) to the bulk action. This term introduces a kinetic contribution for the scalar involving \(\partial_z X_z\), thereby clarifying how the full geometry influences the scalar KK modes together with the existing term. At the same time, this addition preserves gauge symmetry both in the bulk and on the brane just like the Stückelberg mechanism.

It is noteworthy that in a brane model with \(d\) extra dimensions (\(d > 1\)), there exist \(d\) distinct types of scalar KK modes. Interactions among the different types of scalar KK modes induce mixing in the mass matrix of the KK spectrum, leading to oscillation phenomena among these modes. Consequently, although the scalars acquire masses from the geometry of the extra dimensions, the physical states observed on the brane are mass eigenstates arising from this mixing. By diagonalizing the mass matrix, we compute the physical masses of these mixed modes in two illustrative 6D brane models.

The paper is organized as follows: In Sec.~\ref{Section1},  serving as a demonstrative example of the influence of the new term \((\nabla^M X_M)^2\), we investigate the gauge invariance and the mass spectrum of the KK modes in a 5D brane model. Then, in Sec.~\ref{section2}, we extend the study to a 6D brane, where two types of scalar KK modes emerge, leading to interesting oscillation phenomena. Numerical results related to these oscillations are presented in Sec.~\ref{section3}. Finally, a brief summary and discussion are provided in Sec.~\ref{section4}.

\section{Gauge-invariant massive vectors  and their accompanying scalars in 5D brane model}\label{Section1}


We consider a U(1) gauge field coupling with a dilaton field $\pi$ in a  \( (4+1) \)-dimensional braneworld scenario described by a conformal metric with the line element $
 ds^2 = e^{2A (z)}(\hat{g}_{\mu\nu}dx^{\mu}dx^{\nu} +  dz^2)$,  where $z$ denotes the  extra-dimensional coordinate and the warp factor $e^{2A(z)}$depends only on $z$. The action for the bulk U(1) gauge field  is taken to be
\begin{eqnarray}
S_1 
=- \frac{1}{4} \int d^5 x \sqrt{-g}~ 
 e^{\tau \pi}\big(F_{\mu\nu}F^{\mu\nu}+2F_{\mu z}F^{\mu z}+(\nabla_M X^M)^2\big), \label{actionVectorcoup}
	\end{eqnarray}
where  $\tau$ is the coupling constant. This action is invariant under the gauge transformations
\begin{equation}
X_\mu\rightarrow X_\mu+\partial_\mu \Xi,~~
X_z\rightarrow X_z+\partial_z \Xi,
\end{equation}
where $\Xi$  is a bulk scalar field satisfying  the condition 
\begin{equation}\label{xicondition}
\partial_\mu \partial^\mu\Xi+\partial_z^2 \Xi+3(\partial_z \Xi)(\partial_z A)=0.
\end{equation}
	
To derive the effective brane action, we perform a KK decomposition for the bulk field:
\begin{subequations}\label{5Dkkdecomp}
\begin{eqnarray}
  X_\mu &=& \sum_n \hat{X}^{(n)}_\mu(x^\nu) f^{(n)}(z), \label{KK_decomposition_vector} \\
  X_z &=& \sum_m \phi^{(m)}(x^\nu) \rho^{(m)}(z).\label{KK_decomposition_scalar1}
\end{eqnarray}
\end{subequations}
Here, \(\hat{X}^{(n)}_\mu\) represents the \(n\)-level vector KK mode on the brane,  and \(\phi^{(m)}\) represent the  \(m\)-level scalar KK modes on the brane, respectively. It is assumed that the wavefunction profiles $\{f^{(n)}(z)\}$ and $\{\rho^{(m)}(z)\}$ each form a complete set along the extra dimension.  Here we highlight three key consequences within the framework of the new bulk action \eqref{actionVectorcoup}.

\begin{itemize}
\item The gauge invariance of massive vector KK mode
 
 Substituting the KK decompositions \eqref{5Dkkdecomp} into the bulk action, we arrive at the following effective action:
\begin{eqnarray}\label{effective_action_5}
  && S_{\text{eff}} = -\frac{1}{4} \int d^4 x \sqrt{|g|} \sum_{n,m}\big(
   N^{(nm)}\;\hat{F}_{\mu\nu}^{(n)} \hat{F}^{\mu\nu(m)}\nonumber\\
&&+ 2\big[  \hat{X}^{(n)}_\nu  \hat{X}^{\nu(m)} I^{(nm)}
   +  \widetilde{N}^{(nm)}\;\partial_\nu \phi^{(n)}\partial^\nu \phi^{(m)}
   -2\widetilde{I}^{(nm)}  \hat{X}_\nu^{(n)}\partial^\nu \phi^{(m)}
   \big],\nonumber\\
   &&+\big[
(\partial^\mu \hat{X}_\mu^{(n)})(\partial^\mu \hat{X}_\mu^{(m)})N^{(nm)}
+ \phi^{(n)}\phi^{(m)}C^{(nm)}
+2\widetilde{C}^{(nm)}  \partial^\mu \hat{X}^{(n)}_\mu \phi^{(m)}\big],
\end{eqnarray}
where:
  \begin{eqnarray}\label{norm1}
  N^{(nm)}&=&\int e^{A+\tau\pi} dz\;  f^{(n)}  f^{(m)},~~~~~~~~~
  \widetilde{N}^{(nm)} = \int e^{A+\tau\pi} dz\; \rho^{(n)} \rho^{(m)},\\
 I^{(nm)}& = &\int e^{A+\tau\pi} dz\; \partial_z f^{(n)} \partial_z f^{(m)},~~~~
 \widetilde{I}^{(nm)} = \int e^{A+\tau\pi} dz\; \partial_z f^{(n)}  \rho^{(m)},\\
 C^{(nm)}&=&\int  e^{A+\tau\pi} dz\;[\partial_z\rho^{(n)}+3(\partial_z A)\rho^{(n)}][\partial_z\rho^{(m)}+3(\partial_z A)\rho^{(m)}],\\
\widetilde{C}^{(nm)}&=&\int e^{A+\tau\pi} dz\; f^{(n)}[\partial_z\rho^{(m)}+3(\partial_z A)\rho^{(m)}].
  \end{eqnarray} 
The orthonormality conditions, \( N^{(nn)}=1 \) and \( \widetilde{N}^{(mm)}=1 \), serve as  the definitions of the inner product for the basis function sets \( \{ f^{(n)}(z) \} \) and \( \{ \rho^{(m)}(z) \} \), respectively.  Consequently, from the definitions of $\widetilde{I}^{(nm)}$ and $\widetilde{C}^{(nm)}$, we obtain that
\begin{eqnarray}
\partial_z f^{(n)}=\sum_k \widetilde{I}^{(nk)} \rho^{(k)},~~~~~
\partial_z\rho^{(m)}+3(\partial_z A)\rho^{(m)}=\sum_l \widetilde{C}^{(ml)} f^{(l)},
\end{eqnarray}
which lead to
\begin{eqnarray}
I^{(nm)}&=&\sum_{k,l}\widetilde{I}^{(nk)} \widetilde{I}^{(ml)}\widetilde{N}^{(kl)},~~~~~
\widetilde{I}^{(nm)}=\sum_k \widetilde{I}^{(nk)} \widetilde{N}^{(km)},\label{rela5DII}\\
C^{(nm)}&=&\sum_{k,l}\widetilde{C}^{(kn)} \widetilde{C}^{(lm)}N^{(km)},~~~~
\widetilde{C}^{(nm)}=\sum_k \widetilde{C}^{(nk)} N^{(km)}.\label{rela5DCC}
\end{eqnarray}

Then we can rewrite the effective action as
\begin{eqnarray}\label{effective_action_5}
  S_{\text{eff}} = -\frac{1}{4} \sum_{n,m}&&\int d^4 x \sqrt{|g|} \big(
 \widetilde{N}_1^{(nm)}  \hat{F}_{\mu\nu}^{(n)} \hat{F}^{\mu\nu(m)}\nonumber\\
&&+\widetilde{N}_1^{(nm)} 
 \big[\partial_\nu \phi^{(n)}-\sum_k \widetilde{I}^{(kn)} \hat{X}^{(k)}_\nu\big]
 \big[\partial^\nu \phi^{(m)}-\sum_l \widetilde{I}^{(lm)} \hat{X}^{\nu(l)}\big],\nonumber\\
   &&+
N_1^{(nm)}  \big[\partial^\mu \hat{X}_\mu^{(n)}+\sum_{k} \widetilde{C}^{(nk)} \phi^{(k)}\big]
 \big[\partial^\mu \hat{X}_\mu^{(m)}+ \sum_{l} \widetilde{C}^{(ml)} \phi^{(l)}\big].
\end{eqnarray}
which is gauge invariance under the gauge transformations:
  \begin{eqnarray}\label{Gauge_4}
   && \hat{X}_\nu^{(k)} \rightarrow \hat{X}_\nu^{(k)} + \partial_\nu \xi^{(k)},\\
    &&\phi^{(n)} \rightarrow \phi^{(n)} + \sum_k \widetilde{I}^{(kn)} \xi^{(k)},
  \end{eqnarray}
Here $\xi^{n}$ the brane-localized gauge freedom satisfies
\begin{equation}\label{conditiongauge}
\partial^\nu\partial_\nu \xi^{(n)}+\sum_{k,l}(\widetilde{C}^{(nk)}\widetilde{I}^{(lk)} )\xi^{(l)}=0.
\end{equation}
This condition is in fact induced by the constraint \eqref{xicondition} in the bulk. 

The effective action \eqref{effective_action_5} exhibits a structure reminiscent of the Stueckelberg mechanism, yet it possesses distinct physical origins and implications. Unlike the standard Stueckelberg formalism where the gauge parameter is an arbitrary function, the brane-localized gauge freedom $\xi^{(l)}$ here is constrained by the differential condition in Eq. \eqref{conditiongauge}. This restriction implies that while the full gauge redundancy is broken, the a residual gauge symmetry is preserved. 

Consequently, the scalar KK modes cannot be entirely gauged away; they represent genuine physical degrees of freedom rather than mere auxiliary fields. These modes originate intrinsically from the bulk U(1) gauge field, with their effective masses being a direct manifestation of the extra-dimensional geometry. 

\item Eigenstate of vector and scalar KK modes

Starting from the original definitions of \( I^{(nn)} \) and \( C^{(nn)} \), and imposing either Dirichlet or periodic boundary conditions on the basis functions \( f^{(n)} \) and \( \rho^{(m)} \), we introduce \( 2I^{(nn)} = m_v^{2} \) and \( \frac{1}{2}C^{(nn)} = m_{\phi}^{2} \) to derive two constraint equations for the massive vector and scalar KK modes:
\begin{eqnarray}\label{equationsvector}
-e^{-(A+\tau\pi)} \partial_z (e^{(A+\tau\pi)}\partial_z f^{(m)})
&=&m_v^{2} f^{(m)},\\
-e^{2A-\tau\pi}\partial_z [e^{\tau\pi-5A}\partial_z(e^{3A}\rho^{(m)})]
&=&m_{\phi}^{2} \rho^{(m)}.\label{equationsvector}
\end{eqnarray}
After applying the transformation $f^{(n)}=e^{-\frac{\sqrt{3}\tau+1}{2}A}  \bar{f}^{(n)}$ and $\rho^{(m)}=e^{-\frac{3}{2} A} \bar{\rho}^{(m)}$ , the equations reduce to  the following Schrodinger-like forms:
\begin{eqnarray}
(-\partial^2_z+V_{\text{vec}})\bar{f}^{(n)}=m_v^{2} \bar{f}^{(n)},\\
(-\partial^2_z+V_{\text{sca}})\bar{\rho}^{(m)}=m_\phi^{2} \bar{\rho}^{(m)},
\end{eqnarray}
with the effective potentials given by
\begin{eqnarray}
V_{\text{vec}}&=&\frac{(\sqrt{3}\tau+1)^2}{4}A'^2+\frac{\sqrt{3}\tau+1}{2}A'' ,\\
V_{\text{sca}}&=&\frac{(\sqrt{3} \tau-5)^2 }{4}A^{\prime 2}+\frac{\sqrt{3} \tau-5}{2} A^{\prime \prime}.
\end{eqnarray}
Here we identify the dilaton field with $\pi=\sqrt{3}A$ following the literature \cite{Fu2011}, where the brane solution is given as 
\begin{equation}\label{lineelement5D}
A(z)=-\frac{v^2}{9}\big(\ln \cosh ^2(a z)+\frac{1}{2} \tanh ^2(a z)\big).
\end{equation}

We analyze the behavior of the potentials for both vector and scalar KK modes in the limits $z\rightarrow 0$ and $z \rightarrow \infty$, obtaining:
\begin{eqnarray}
V_{\text{vec}}(z\rightarrow 0) &\rightarrow& -\frac{1}{6} a^2 v^2(1+\sqrt{3} \tau),~~
V_{\text{vec}}(z\rightarrow \infty)\rightarrow \frac{1}{81} a^2v^4(1+\sqrt{3} \tau)^2,\\
V_{\text{sca}}(z\rightarrow 0) &\rightarrow & -\frac{1}{6} a^2 v^2(\sqrt{3} \tau-5),~~
V_{\text{sca}}(z\rightarrow \infty)\rightarrow \frac{1}{81} a^2v^4(\sqrt{3} \tau-5)^2.
\end{eqnarray}
It is evident that for $\tau>-1/\sqrt{3}$, a zero vector KK mode is always localized, while a massless scalar mode only appears when $\tau>5/\sqrt{3}$. When $\tau$ is sufficiently large, both massive vector and scalar modes exist. The corresponding effective potentials are illustrated in Fig.~\ref{fig5Dpoten}. Moreover, as shown in Table~\ref{5Dmasscomp}, the masses of the scalar modes are smaller than those of the vector modes.
\begin{figure*}[htbp]
\centering
      \includegraphics[width=0.9\textwidth]{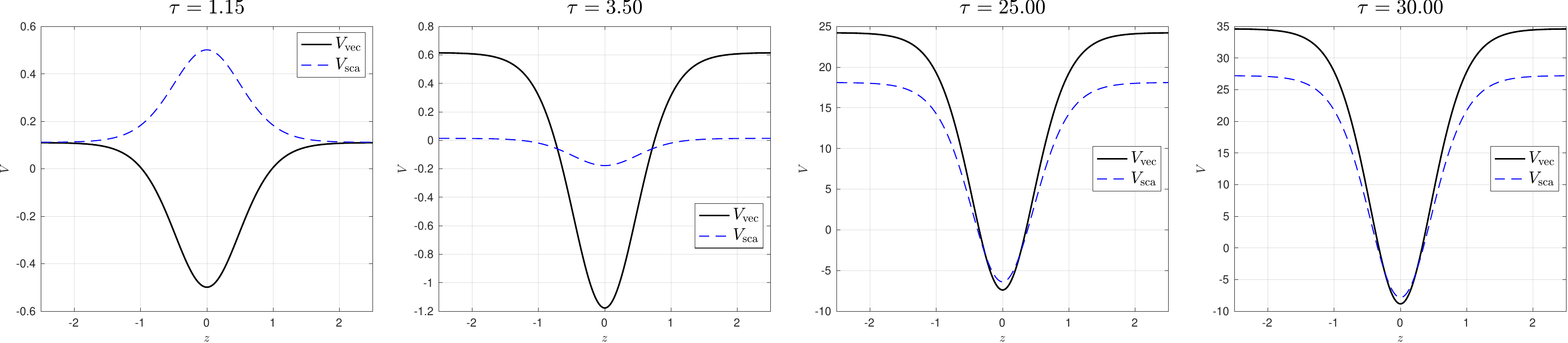} 
  \caption{The effective potentials for the vector and scalar KK modes with $a=1,v=1$.  }
  \label{fig5Dpoten}
\end{figure*}
\begin{table}[htbp]
\centering
\begin{NiceTabular}{c |cc| cc|  cc|cc}
\Hline
\Block{2-1}{$n_2$} & \Block{1-2}{$\tau=1.15$} & & \Block{1-2}{$\tau=3.5$} & & \Block{1-2}{$\tau=30$} &  & \Block{1-2}{$\tau=40$}\\
\Hline
& $m_v^{2}$ & $m_{\phi}^{2}$  & $m_v^{2}$ & $m_{\phi}^{2}$ & $m_v^{2}$ &$m_{\phi}^{2}$ & $m_v^{2}$ &$m_{\phi}^{2}$\\
\Hline
0 & 0 &  & 0 & 0  & 0          & 0         & 0         &  0 \\
1 &    &   &   &     &15.59     & 13.59   & 21.38   & 19.38\\
2 &    &  &   &     & 26.89    & 22.82   & 38.56   &34.53\\
3 &    &  &   &      &33.41    & 27.09   &  51.24   &45.11\\
4 &    &  &   &      &            &             &  58.96   &50.54 \\
\Hline
\Hline
\end{NiceTabular}
\caption{Squared mass eigenvalues for vector and scalar KK modes.}
\label{5Dmasscomp}
\end{table}

\end{itemize}

\section{Gauge-invariant massive vectors  and their accompanying scalars in 6D brane model}\label{section2}

We consider a massless U(1) gauge field in a  \( (4+2) \)-dimensional brane world. The line element  for the brane is given by:
\begin{eqnarray}\label{unconformalmetric}
 ds^2 = e^{2A
 (z,y)}\hat{g}_{\mu\nu}dx^{\mu}dx^{\nu} + e^{2B_1(z,y)} dz^2
  +e^{2B_2(z,y)} dy^2,
\end{eqnarray}
where the warp factors $e^{2A(z,y)}$ and $e^{2B_2(z,y)},e^{2B_2(z,y)}$ are functions of the two extra dimensions. 

\subsection{Review the gauge invariance of massive vector KK modes }

For a free bulk U(1) gauge field in the 6D brane world, after performing the KK decomposition 
\begin{eqnarray}\label{kkdecomp}
  X_\mu = \sum_n \hat{X}^{(n)}_\mu(x^\nu) f^{(n)}(z,y), 
  X_z = \sum_n \phi^{(n)}(x^\nu) \rho^{(n)}(z,y), 
  X_y = \sum_n \varphi^{(n)}(x^\nu) \chi^{(n)}(z,y), 
\end{eqnarray}
where \(\hat{X}^{(n)}_\mu\) denotes \(n\)-level vector KK mode, and \(\phi^{(m)}, \varphi^{(m)}\) are two types of  scalar ones, the effective action takes the form
\begin{eqnarray}\label{effective_action_6}
  S_{\text{eff1}} = -\frac{1}{4} &&\sum_{n,m}\int d^4 x \sqrt{|g|} \big(
  N^{(nm)}\;\hat{F}_{\mu\nu}^{(n)} \hat{F}^{\mu\nu(m)}\nonumber\\
 &&+ 2\big[ \hat{X}^{(n)}_\nu \hat{X}^{\nu(m)} I_1^{(nm)}
 -2\hat{X}_\nu^{(n)}\partial^\nu \phi^{(m)}\widetilde{I}_1^{(nm)}
   + \widetilde{N}_1^{(nm)}\;\partial_\nu \phi^{(n)}\partial^\nu \phi^{(m)}\big]\nonumber\\
 &&+ 2\big[ \hat{X}^{(n)}_\nu \hat{X}^{\nu(m)} I_2^{(nm)}
 -2\hat{X}_\nu^{(n)}\partial^\nu \varphi^{(m)}\widetilde{I}_2^{(nm)}
   + \widetilde{N}_2^{(nm)}\;\partial_\nu \varphi^{(n)}\partial^\nu \varphi^{(m)}\big]\nonumber\\
 &&+2\big[C_1^{(nm)}\phi^{(n)}\phi^{(m)}+C_2^{(nm)}\varphi^{(n)}\varphi^{(m)}-2\widetilde{C}^{(nm)}\phi^{(n)}\varphi^{(m)}\big]
   \big)
\end{eqnarray}
where
  \begin{eqnarray}\label{norm1}
  N^{(nm)}&=&\int e^{B_1+B_2} dzdy\;  f^{(n)}  f^{(m)},\\
 \widetilde{N}_1^{(nm)} &=& \int e^{2A-B_1+B_2} dzdy\; \rho^{(n)} \rho^{(m)},~~~~~~~~\label{norm2}
  \widetilde{N}_2^{(nm)} = \int e^{2A+B_1-B_2} dzdy\; \chi^{(n)} \chi^{(m)},\\\label{norm3}
 I_1^{(nm)}& = &\int e^{2A-B_1+B_2} dzdy\; \partial_z f^{(n)} \partial_z f^{(m)},~~
 \widetilde{I}_1^{(nm)} = \int e^{2A-B_1+B_2} dzdy\; \partial_z f^{(n)}  \rho^{(m)},\\
 I_2^{(nm)}& = &\int e^{2A+B_1-B_2} dzdy\; \partial_y f^{(n)} \partial_y f^{(m)},~~
 \widetilde{I}_2^{(nm)} = \int e^{2A+B_1-B_2} dzdy\; \partial_y f^{(n)}  \chi^{(m)},\\
 C_1^{(mm)}& = &\int e^{4A-B_1-B_2} dzdy\; \partial_y \rho^{(m)} \partial_y \rho^{(m)} ,~~
 C_2^{(kk)}= \int e^{4A-B_1-B_2} dzdy\; \partial_z \chi^{(k)} \partial_z \chi^{(k)},\\
 \widetilde{C}^{(mk)}&=&\int e^{4A-B_1-B_2} dzdy\; \partial_y \rho^{(m)} \partial_z \chi^{(k)}.
  \end{eqnarray}
We note that the scalar KK modes  $\phi^{(m)}$ and $\varphi^{(k)}$ mix with each other through the term $F_{yz}F^{yz}$ in the 6D bulk action. At the fist sight, the presence of these scalar mass and mixing terms appears to make gauge invariance difficult to maintain. However, as we show below, the effective actions remains gauge invariant due to nontrivial relations among the vector-scalar and scalar-scalar couplings.

Although there are two types of scalar KK modes, they couple to the vector KK modes in the same manner. Therefore, an analogous relation holds between $I_1^{(nn)}$ ($I_2^{(nm)}$) and $\widetilde{I}_1^{(nm)}$ ($\widetilde{I}_2^{(nm)}$) as specified in \eqref{rela5DII}. While by defining two auxiliary matrices
\begin{eqnarray}
T_1^{(nm)}&=&\int e^{B_1+B_2}dzdy \;e^{2A-B_1-B_2}\partial_y \rho^{(m)} f^{(n)},\\
T_2^{(nk)}&=&\int e^{B_1+B_2}dzdy \; e^{2A-B_1-B_2}\partial_z \chi^{(k)} f^{(n)},
\end{eqnarray}
 together with the completeness of the basis functions, we obtain other relations:
\begin{eqnarray}
C_1^{(nm)}=\sum_{k,l} T_1^{(kn)}T_1^{(lm)}N^{(kl)},
C_2^{(nm)}=\sum_{k,l} T_2^{(kn)}T_2^{(lm)}N^{(kl)},
\widetilde{C}^{(nm)}&=&\sum_{k,l} T_1^{(kn)}T_2^{(lm)}N^{(kl)}.\nonumber
\end{eqnarray}
With these identities, the effective action can be rewritten as 
\begin{eqnarray}
S^{(4)} = -\frac{1}{4} &&\sum_{n,m}\int d^4 x \sqrt{|g|} \big(
  \hat{F}_{\mu\nu}^{(n)} \hat{F}^{\mu\nu(n)}\nonumber\\
&&+ \widetilde{N}_1^{(nm)} 
 \big[\partial_\nu \phi^{(n)}-\sum_k \widetilde{I}_1^{(kn)} \hat{X}^{(k)}_\nu\big]
 \big[\partial^\nu \phi^{(m)}-\sum_l \widetilde{I}_1^{(lm)} \hat{X}^{\nu(l)}\big]\nonumber\\
&&+\widetilde{N}_2^{(nm)} 
 \big[\partial_\nu \varphi^{(n)}-\sum_k \widetilde{I}_2^{(kn)} \hat{X}^{(k)}_\nu\big]
 \big[\partial^\nu \varphi^{(m)}-\sum_l \widetilde{I}_2^{(lm)} \hat{X}^{\nu(l)}\big]\nonumber\\
&&+\sum_{k,l} N^{(kl)}
\big[T_1^{(kn)}\phi^{(n)}-T_2^{(kn)}\varphi^{(n)}\big]
\big[T_1^{(lm)}\phi^{(m)}-T_2^{(lm)}\varphi^{(m)}\big] .
\end{eqnarray}
We observe that under the gauge transformations
\begin{eqnarray}\label{Gauge_4}
\hat{X}_\nu^{(k)} \rightarrow \hat{X}_\nu^{(k)} + \partial_\nu \xi^{(k)},
\phi^{(n)} \rightarrow \phi^{(n)} +  \sum_k\widetilde{I}_1^{(kn)} \xi^{(k)},
\varphi^{(n)} \rightarrow \varphi^{(n)} +  \sum_k\widetilde{I}_2^{(ln)} \xi^{(l)}
\end{eqnarray}
the first three terms remain gauge invariant. But an additional term of the form $\sum_{k,l} (T_1^{(kl)} \widetilde{I}_1^{(nl)} - T_2^{(kl)} \widetilde{I}_2^{(nl)}\xi^{(n)}$ emerges in the final expression. Utilizing the symmetry of mixed partial derivatives $\partial_y \partial_z f^{(n)} = \partial_z \partial_y f^{(n)}$, we obtain
\begin{equation}
\sum_{l}\left( T_1^{(kl)}\widetilde{I}_1^{(nl)} - T_2^{(kl)}\widetilde{I}_2^{(nl)} \right)=0.
\end{equation}
This term thus vanishes. Consequently, the effective action is gauge-invariant, consistent with the result established in Ref.\cite{Fu2025}. 

However, when we examine the equations that govern the KK modes under Dirichlet or periodic boundary conditions on the basis functions 
\begin{eqnarray}
-e^{-(B_1+B_2)} \big(\partial_z (e^{2A-B_1+B_2}\partial_z f^{(m)})+  \partial_y( e^{2A+B_1-B_2}\partial_y f^{(m)})\big)&=&m_v^{2} f^{(m)},\\
-e^{-(2A-B_1+B_2)} \partial_y (e^{4A-B_1-B_2} \partial_y \rho^{(m)} )&=&m_\phi^2 \rho^{(m)}, \label{sca1}\\
-e^{-(2A+B_1-B_2)} \partial_z (e^{4A-B_1-B_2} \partial_z \chi^{(k)} )&=&m_\varphi^2 \chi^{(k)},\label{sca2}
\end{eqnarray}
we note that the vector fields acquire masses from the full extra spacetime, whereas the scalar fields receive mass contributions from only one of the extra dimensions, either $y$ or $z$. Because both the warp factor
 and the wave functions depend on $y$ and $z$, the equations for the scalar modes cannot be solved independently. Therefore, as in the five-dimensional case, it is necessary to modify the bulk action by including the term $(\nabla_M X^M)^2$.

\subsection{New bulk action for the U(1) gauge field and their KK modes}

By the KK decomposition,  the additional term $(\nabla_M X^M)^2$ becomes
\begin{eqnarray}\label{6Dadd1}
&&\int dzdy\sqrt{-g}(\nabla_M X^M)^2=\sum_{n,m}\sqrt{-\hat{g}}\big(
N^{(nm)}(\partial^\mu \hat{X}_\mu^{(n)}) (\partial^\mu \hat{X}_\mu^{(m)})
+C_1^{'(nm)}\phi^{(n)} \phi^{(m)}
+C_2^{'(nm)} \varphi^{(n)} \varphi^{(m)}\nonumber\\
&&~~~~~~~~~~~~~~~~~~~~
+2\widetilde{I}_1^{'(nm)}\partial^\mu \hat{X}^{(n)}_\mu \phi^{(m)}
+2\widetilde{I}_2^{'(nm}\partial^\mu \hat{X}^{(n)}_\mu \varphi^{(m)}
+2\widetilde{I}^{'(nm)} \phi^{(n)} \varphi^{(m)}
\big),
\end{eqnarray}
where 
\begin{eqnarray}
C_1^{'(nm)}&=&\int {e}^{-(4 A+B_1+B_2)} dzdy \;(\partial_z({e}^{4 A-B_1+B_2} \rho^{(n)}))(\partial_z({e}^{4 A-B_1+B_2} \rho^{(m)})),\\
C_2^{'(nm)}&=&\int {e}^{-(4 A+B_1+B_2)} dzdy \;(\partial_y({e}^{4 A+B_1-B_2}\chi^{(n)}))(\partial_y({e}^{4 A+B_1-B_2}\chi^{(m)})),\\
\widetilde{I}_1^{'(nm)}&=&\int e^{-2A}dzdy \;f^{(n)}(\partial_z({e}^{4 A-B_1+B_2} \rho^{(m)})),\\
\widetilde{I}_2^{'(nm)}&=&\int e^{-2A}dzdy \;f^{(n)}(\partial_y({e}^{4 A+B_1-B_2}\chi^{(m)})),\\
\widetilde{C}^{'(nm)}&=& \int {e}^{-(4 A+B_1+B_2)} dzdy \;(\partial_z({e}^{4 A-B_1+B_2} \rho^{(n)}))(\partial_y({e}^{4 A+B_1-B_2}\chi^{(m)})).
\end{eqnarray}
Considering 
\begin{eqnarray}
{e}^{-2A-B_1-B_2}(\partial_z({e}^{4 A-B_1+B_2} \rho^{(m)}))&=&\sum_n \widetilde{I}_1^{'(nm)} f^{(n)} ,\\
{e}^{-2A-B_1-B_2}(\partial_y({e}^{4 A+B_1-B_2} \chi^{(k)}))&=&\sum_n \widetilde{I}_2^{'(nk)} f^{(n)} ,
\end{eqnarray}
we get
\begin{eqnarray}
C_1^{'(nm)}=\sum_{k,l}\widetilde{I}_1^{'(kn)}\widetilde{I}_1^{'(lm)}N^{(kl)},
C_2^{'(nm)}=\sum_{k,l}\widetilde{I}_2^{'(kn)}\widetilde{I}_2^{'(lm)}N^{(kl)},
\widetilde{I}^{'(nm)}= \sum_{k,l}\widetilde{I}_1^{'(kn)}\widetilde{I}_2^{'(lm)}N^{(kl)}.\nonumber
\end{eqnarray}
This makes the effective term \eqref{6Dadd1} for $(\nabla_M X^M)^2$ at last turn to
\begin{eqnarray}
\sum_{n}\sqrt{-\hat{g}}\big[\partial^\mu \hat{X}_\mu^{(n)}
+\sum_k \widetilde{I}_1^{'(nk)}\;\phi^{(k)}
+\sum_k  \widetilde{I}_2^{'(nk)}\;\varphi^{(k)}\big]^2,
\end{eqnarray}
which is gauge invariance under the gauge transformations \eqref{Gauge_4} with the condition
\begin{equation}
\partial^\mu\partial_\mu \xi^{(n)}
+\sum_{k,l}(\widetilde{I}_1^{'(nk)}\widetilde{I}_1^{(lk)}+\widetilde{I}_2^{'(nk)}\widetilde{I}_2^{(lk)})\xi^{(l)}=0.
\end{equation}

Now for the scalars, the masses of them are $C_1^{(nm)}+C_1^{'(nm)}=m_{\phi}^{2}\delta^{nm}$ and $C_2^{(nm)}+C_2^{'(nm)}=m_{\varphi}^{2}\delta^{nm}$, then there are 
\begin{eqnarray}
&&-e^{-(2A-B_1+B_2)} \partial_y (e^{4A-B_1-B_2} \partial_y \rho^{(m)} )
-{e}^{2 A}\partial_z\big( {e}^{-(4 A+B_1+B_2)}\partial_z({e}^{4 A-B_1+B_2} \rho^{(n)})\big)=m_\phi^2 \rho^{(m)}, \nonumber\\
\label{lsca1}\\
&&-e^{-(2A+B_1-B_2)} \partial_z (e^{4A-B_1-B_2} \partial_z \chi^{(k)} )
-{e}^{2 A}\partial_y\big( {e}^{-(4 A+B_1+B_2)}\partial_y({e}^{4 A+B_1-B_2}\chi^{(k)})\big) =m_\varphi^2 \chi^{(k)},\nonumber\\
\label{lsca2}
\end{eqnarray}
The eigenfunctions for these three equations \eqref{equationsvector}, \eqref{lsca1} and \eqref{lsca2} are chosen as the basis functions $f^{(n)}$, $\rho^{(n)}$ and $\chi^{(n)}$. It is noted that these two types of scalars interact with each other and can exhibit mixing.

We note that in most 6D brane models, the warp factors are assumed to depend only on a single extra dimension $y$. By separating the three basis functions as $f^n = R_1^{(n,l)}(r) \Theta^{(l)}(\theta)$, $\rho^m = R_2^{(m,l)}(r) \Theta^{(l)}(\theta)$, and $\chi^k = R_3^{(k,l)}(r) \Theta^{(l)}(\theta)$,  assuming
$\partial_\theta^2 \Theta^{(l)} + l^2 \Theta^{(l)} = 0$,
and applying the transformations
$d\bar{r}=e^{-A+B_1}dr$,
$\bar{R}_1^{(n,l)}=e^{\frac{1}{2}(A+B_2)}R_1^{(n,l)}$,
$\bar{R}_2^{(m,l)}=e^{\frac{1}{2}(3A-2B_1+B_2)}R_2^{(m,l)}$,
$\bar{R}_3^{(k,l)}=e^{\frac{1}{2}(3A-B_2)}R_3^{(k,l)}$,
we find that the three equations \eqref{equationsvector}, \eqref{lsca1}, and \eqref{lsca2} reduce to Schrodinger-like equations:
\begin{eqnarray}
&&-\partial{\bar{r}}^2 \bar{R}_1^{(n,l)} + V_{\text{eff1}} \bar{R}_1^{(n,l)} = m_v^{2} \bar{R}_1^{(n,l)}, \label{sch1}\\
&&-\partial{\bar{r}}^2 \bar{R}_2^{(m,l)} + V_{\text{eff2}}\bar{R}_2^{(m,l)} = m\phi^2 \bar{R}_2^{(m,l)}, \label{sch2}\\
&& -\partial{\bar{r}}^2 \bar{R}_3^{(k,l)} + V_{\text{eff3}}\bar{R}_3^{(k,l)} =m\varphi^2 \bar{R}_3^{(k,l)}. \label{sch3}
\end{eqnarray}
Here, the effective potentials are given by
\begin{eqnarray}
V_{\text{eff1}}&=&\frac{1}{4}(\partial{\bar{r}} (A+B_2))^2+\frac{1}{2}\partial{\bar{r}}^2(A+B_2)+ e^{2A-2B_2}l^2,\\
V_{\text{eff2}}&=&\frac{1}{4}(\partial_{\bar{r}} (3A-2B_1+B_2))^2+\frac{1}{2}\partial_{\bar{r}}^2(3A-2B_1+B_2)+ e^{2A-2B_2}l^2\nonumber\\
&&-\left[\partial_{\bar{r}}^2( 4A-B_1+B_2)+\partial_{\bar{r}}(-A-B_1)\partial_{\bar{r}}( 4A-B_1+B_2)\right],\\
V_{\text{eff3}}&=&\frac{1}{4}(\partial_{\bar{r}} (3A-B_2))^2+\frac{1}{2}\partial_{\bar{r}}^2(3A-B_2)+ e^{2A-2B_2}l^2.
\end{eqnarray}
These effective potentials can then be analyzed to study the mass spectra of the Kaluza–Klein modes in specific brane models.

\section{ The mixing and oscillations of the two scalar KK modes }\label{section3}

It is noted that the two types of the scalars couple together, for which the coupling coefficients $C^{(mk)}$ read
\begin{equation}
C^{(mk)}=\widetilde{C}^{(mk)}+\widetilde{C}^{'(mk)}=\int d\bar{r}\; e^{-2B_2}\partial_{\bar{r}}(e^{A+B_2} \bar{R}_2^{(m,l)}\bar{R}_3^{(k,l)}).
\end{equation}
These couplings  lead to the oscillation of the corresponding scalar KK modes.  Typically, the physical masses $\mathcal{M}^{2(n)}_A$ of these oscillations are obtained by diagonalizing the mass matrix \( \mathcal{M}^2 \) , which is made by:
\begin{equation}
\mathcal{M}^2=\left(\begin{array}{ll}
{\left[m_\phi^2\right]_{n \times n}} & {[C]_{n \times n}} \\
{\left[C^T\right]_{n \times n}} & {\left[m_{\varphi}^2\right]_{n \times n}}
\end{array}\right)
\end{equation}
Here, the diagonal blocks correspond to the KK mass-squared of the two types of scalar fields, while the off-diagonal blocks characterize the coupling between them.


In the following, we will calculate the masses of these oscillations in two different 6D brane models.

\subsection{Case 1}
In Ref.~\cite{Antoniadis1998} some 6D thick-brane models from a cosmological perspective, where the metric is assumed as:
\begin{equation}\label{line-element1}
d s^2=M^2(r) g_{\mu \nu} d x^\mu d x^\nu+d r^2+L(r)^2 d \theta^2
\end{equation}
The metric functions $M(r)$ and $L(r)$ depend only on the radial coordinate $r$, with a periodic angular coordinate $\theta \in[0,2 \pi]$. To solve equations for the vector and scalar KK mode, we consider one of the solutions :
\begin{equation}
M(r)=\cosh (a r), \quad L(r)=\Re a \sinh (a r),
\end{equation}
where $a$ is related to the Hubble constant $H_0$, and $\Re$ depends on the brane tension. For convenience, we set $\Re=1$.

\begin{figure*}[!htbp]
\centering
\subfigure[$l=0$]{
\includegraphics[width=0.3\textwidth]{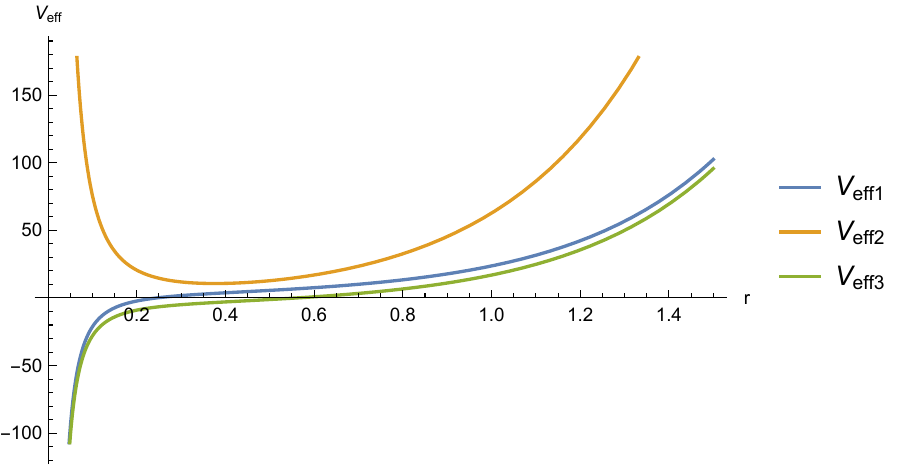} }
\subfigure[$l=1$]{
\includegraphics[width=0.3\textwidth]{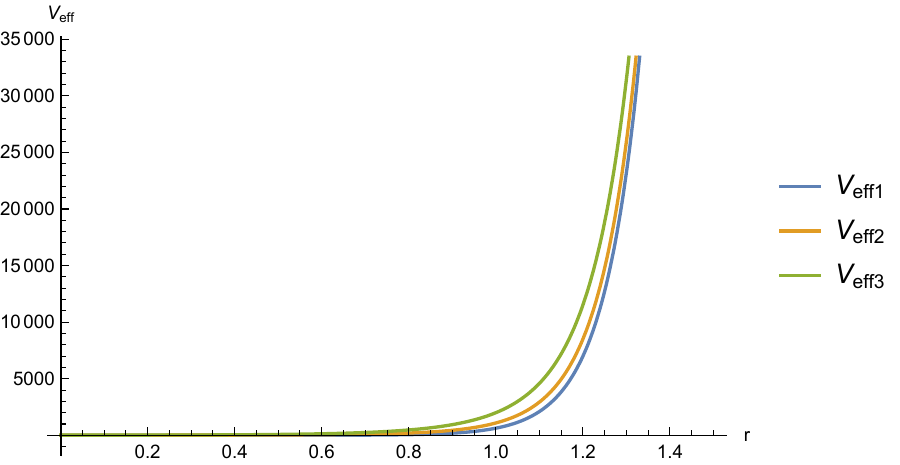} }
\subfigure[$l=2$]{
\includegraphics[width=0.3\textwidth]{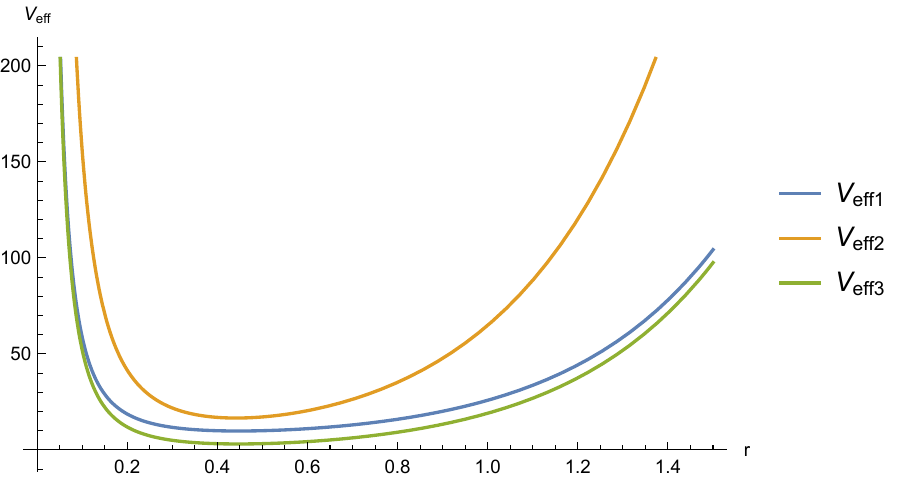} }
\hfill
\caption{Effective potentials for the KK modes in a 6D brane with line element \eqref{line-element1}.}
\label{case1_potentials}
\end{figure*}

Comparing the notation of the metric in \eqref{unconformalmetric}, we obtain:
\begin{equation}
e^{2 A}=M^2, \quad e^{2 B_1}=1, \quad e^{2 B_2}=L^2
\end{equation}
where $L(r)>0$ for $r>0$; otherwise, $e^B=|L|$. The corresponding effective potentials for the Kaluza–Klein (KK) modes are plotted in Figure \ref{case1_potentials}.
We also compute the mass spectra for these modes, as summarized in Table~\ref{6Dmasscomp}.
\begin{table}[htbp]
\centering
\scalebox{0.8}{
\begin{NiceTabular}{c |ccc| ccc|  ccc|ccc}
\Hline
\Block{2-1}{n} & \Block{1-3}{$l=0$} & & & \Block{1-3}{$l=1$} & & & \Block{1-3}{$l=2$} & & & \Block{1-3}{$l=3$} & & \\
\Hline
& $m_v^{2}$ & $m_{\phi}^{2}$ & $m_{\varphi}^{2}$ 
& $m_v^{2}$ & $m_{\phi}^{2}$ & $m_{\varphi}^{2}$ 
& $m_v^{2}$ & $m_{\phi}^{2}$ & $m_{\varphi}^{2}$
& $m_v^{2}$ & $m_{\phi}^{2}$ & $m_{\varphi}^{2}$ \\
\Hline
0 & 13.90   & 31.77   & 22.73    &19.21 & 33.70  &24.24     & 25.49 &38.88 &28.33               & 32.75 & 46.31  & 34.27  \\
1 & 45.96  & 81.81 & 63.67      & 54.99  & 84.59 & 66.02    & 65.37 & 91.95 &72.31             & 76.67 & 102.35  & 81.23\\
2 & 96.23  & 150.1 & 122.81     & 109.06 & 153.71  &126.02    & 123.44 & 163.24 & 134.48   & 138.78 & 176.59 &  146.37\\
3 & 164.73  & 236.62 &200.18   &181.34 & 241.07 &204.23     & 199.7 &252.73 & 214.85    &219.07  &269.02  &  229.70 \\
\Hline
\Hline
\end{NiceTabular}
}
\caption{Mass spectra for the KK modes in a 6D brane with line element \eqref{line-element1}.}
\label{6Dmasscomp}
\end{table}

In this brane mode, the corresponding coupling matrix for different $l$ is provided in Table~\ref{6DssMatrix}. 
\begin{table}[htbp]
\centering
\begin{NiceTabular}{c |c| c| c| c}
\Hline
 & $l=0$ & $l=1$ & $l=2$ & $l=3$ \\
\Hline
$C^{(nm)}$ &  
\scalebox{0.8}{\footnotesize$\begin{pmatrix}
9.54 & 7.20 & 6.22 & 5.58 \\
4.09 & 14.03 & 11.93 & 10.85 \\
5.11 & 7.90 & 18.74 & 16.70 \\
3.71 & 9.64 & 12.27 & 23.83
\end{pmatrix}$} &
\scalebox{0.8}{\footnotesize$\begin{pmatrix}
8.95 & 6.40 & 5.21 & 4.47 \\
3.30 & 12.73 & 10.34 & 8.87 \\
4.33 & 6.20 & 16.39 & 14.04 \\
2.73 & 7.94 & 9.25 & 19.97
\end{pmatrix}$} &
\scalebox{0.8}{\footnotesize$\begin{pmatrix}
6.39 & 3.07 & 1.89 & 1.28 \\
0.29 & 8.25 & 5.05 & 3.46 \\
1.90 & 1.26 & 10.10 & 6.92 \\
0.19 & 3.49 & 2.50 & 11.90
\end{pmatrix}$} &
\scalebox{0.8}{\footnotesize$\begin{pmatrix}
7.27 & 4.17 & 2.94 & 2.23 \\
1.35 & 9.88 & 6.81 & 5.27 \\
2.58 & 3.14 & 12.51 & 9.35 \\
1.01 & 4.83 & 5.26 & 15.19
\end{pmatrix}$} \\
\Hline
\end{NiceTabular}
\caption{Coupling coefficients of scalar-scalar couplings in a 6D brane with line element \eqref{line-element1}.}
\label{6DssMatrix}
\end{table}
We can calculate the masses of these modes by diagonalizing the mass matrix \( \mathcal{M}^2 \), which is shown in Table. \ref{physmass}.
\begin{table}[htbp]
\centering
\scalebox{0.8}{
\begin{NiceTabular}{c|cc|cc|cc|cc}
\Hline
\Block{2-1}{n} & \multicolumn{2}{c|}{$l=0$} & \multicolumn{2}{c|}{$l=1$} & \multicolumn{2}{c|}{$l=2$} & \multicolumn{2}{c}{$l=3$} \\
\Hline
& $m_v^{2}$ & $\mathcal{M}^{2(n)}_A$ & $m_v^{2}$ & $\mathcal{M}^{2(n)}_A$ & $m_v^{2}$ & $\mathcal{M}^{2(n)}_A$ & $m_v^{2}$ & $\mathcal{M}^{2(n)}_A$ \\
\Hline
0 & 13.90 & 15.33        & 19.21 & 18.00          & 25.49 & 25.24             & 32.75 & 30.67 \\
1 & 45.96 & 36.00        & 54.99 & 37.70         & 65.37 & 41.59            & 76.67 & 49.18 \\
2 & 96.23 & 55.83        & 109.06 & 59.61      & 123.44 & 69.45          & 138.78 & 77.48 \\
3 & 164.73 & 86.00    & 181.34 & 88.48       & 199.7 & 94.31            & 219.07 & 105.13 \\
\Hline
\end{NiceTabular}
}
\caption{Mass spectra for the vector and physical scalar KK modes in a 6D brane with line element \eqref{line-element1}.}
\label{physmass}
\end{table}
It suggests that the vector KK modes are heavier than the scalars except for $l=0, n=0$.

\subsection{Case 2}

Let us consider another 6D brane model presented in Ref. \cite{Carroll:2003}, which features football‑shaped extra dimensions. The line element is assumed to be  
\begin{equation}\label{line-element2}
ds^2=\eta_{\mu\nu}dx^\mu dx^\nu+\psi(r)(dr^2+r^2d\phi^2).
\end{equation}
The solution for $\psi(r)$ is $\frac{4 \alpha^2 a_0^2}{r^2\left[(r / r_0)^\alpha+(r / r_0)^{-\alpha}\right]^2}$, where $r_0$ is an arbitrary parameter, while $\alpha$ and $a_0$ are observable quantities related to the brane tension.  

Comparing this with the line element used in the present work, we obtain  
\begin{eqnarray}
A=0,~
B_1=\log\left(\frac{4 \alpha^2 a_0^2}{r^2\left[(r / r_0)^\alpha+(r / r_0)^{-\alpha}\right]^2}\right),
~
B_2=\log\left(\frac{4 \alpha^2 a_0^2}{\left[(r / r_0)^\alpha+(r / r_0)^{-\alpha}\right]^2}\right).
\end{eqnarray}
Interestingly, in this case the effective potentials $V_{\text{eff2}}$ and $V_{\text{eff3}}$ coincide. Consequently, the mass spectra for the two types of scalar KK modes are identical. Following the same procedure with case 1, we can calculate the physical masses of the scalar KK modes arising from their interactions; the results are presented in Table \ref{case2masscom}.

\begin{table}[htbp]
\centering
\scalebox{0.8}{
\begin{NiceTabular}{c|cc|cc|cc|cc}
\Hline
\Block{2-1}{$n$} & \multicolumn{2}{c|}{$l=0$} & \multicolumn{2}{c|}{$l=1$} & \multicolumn{2}{c|}{$l=2$} & \multicolumn{2}{c}{$l=3$} \\
\Hline
& $m_v^{2}$ & $\mathcal{M}^{2(n)}_A$ & $m_v^{2}$ & $\mathcal{M}^{2(n)}_A$ & $m_v^{2}$ & $\mathcal{M}^{2(n)}_A$ & $m_v^{2}$ & $\mathcal{M}^{2(n)}_A$ \\
\Hline
0 & 1.12 & 1.05       & 8.25 & 10.85         & 24.06 & 28.03            & 48.06 & 52.28 \\
1 & 12.36 & 4.21        & 25.28 & 11.33         & 48.33 & 28.04           & 80.25 & 52.28 \\
2 & 32.95 & 19.25        & 50.98 & 30.34      & 80.88 & 53.94        & 120.54 & 85.70\\
3 & 62.80 & 22.84      & 85.79 & 31.45      & 121.90 & 54.01          & 168.96 & 85.71 \\
\Hline
\end{NiceTabular}
}
\caption{Mass spectra for the vector and physical scalar KK modes in the 6D brane with the line element \eqref{line-element2}.}
\label{case2masscom}
\end{table}
We observe that for $n\neq 0$, the vector modes remain heavier than the scalar modes. For larger values of $l$, some degenerate states of the scalars appear.

\section{Summary and discussion}\label{section4}

In this work, by incorporating a \((\nabla^M X_M)^2\) term for a bulk \(U(1)\) gauge field, we derived a Stückelberg-like effective action for the massive Kaluza-Klein (KK) modes of the gauge field. Gauge invariance of the effective action is ensured by an imposed condition on the brane-localized gauge freedom. This construction deviates from the conventional Stückelberg mechanism in two critical aspects: first, the scalar KK modes are not introduced ad hoc but emerge naturally from the extra-dimensional components of the bulk \(U(1)\) gauge field; second, the scalar KK modes can acquire masses from the entire geometric structure. Notably, when the brane possesses more than one extra dimension, additional scalar modes arise.

We first examined the scenario within a five-dimensional brane-world framework. Numerical analyses of this model reveal that the massive vector KK modes are heavier than the scalar KK modes. Extending this analysis to six-dimensional brane models, two distinct types of scalar KK modes emerge. We focused specifically on the scalar–scalar mixed couplings induced by the geometric background. These mixed couplings give rise to mass oscillations among the scalar modes; by diagonalizing the mass matrix, we obtained the physical masses of the scalar sector.

In summary, the scalar KK modes discussed herein are distinguishable from both conventional Stückelberg scalars and Goldstone bosons, attributable to their unique coupling structure with vector modes and among themselves. For future work, it would be intriguing to further explore the properties of these scalar modes and their mixed couplings within the framework of quantum field theory, with particular emphasis on their phenomenological implications.


\providecommand{\href}[2]{#2}\begingroup\raggedright\endgroup

\end{document}